\documentclass[aps,amsmath,amssymb]{revtex4}
\usepackage{graphicx}
\usepackage{dcolumn}
\usepackage{bm}
\usepackage{hyperref}
\usepackage{xcolor}
\usepackage{epsfig}
\usepackage{multirow}
\usepackage{booktabs}
\usepackage{cancel}
\usepackage{etoolbox}
\usepackage{tikz}
\usepackage{orcidlink}

\begin{document}
	\title{Statistical Properties of the $q$-Deformed Monolayer and Bilayer Graphene}
	
	\author{\orcidlink{0000-0002-0260-0910}Omar Pedraza$^{*,1}$, \orcidlink{0000-0002-2180-3895}Erik Díaz-Bautista$^{\dagger,2}$, \orcidlink{0000-0002-6395-6777}L. A. López$^{*,3}$ and  \orcidlink{0000-0003-0321-3442}Daniel O-Campa$^{\dagger\ddagger,4}$\\
		$^{*}$Área Académica de Matemáticas y Física, UAEH, 
		Carretera Pachuca-Tulancingo Km. 4.5,\\ C. P. 42184, Mineral de la Reforma, Hidalgo, México.\\
        $^{\dagger}$Escuela Superior de Física y Matemáticas del Instituto Politécnico Nacional, 07738 Ciudad de México, México.\\
        $^{\ddagger}$División de Ingeniería Electromecánica, Instituto Tecnológico Superior del Occidente del Estado de Hidalgo, 42700, Mixquiahuala de Juárez, Hidalgo, México.\\
		e-mail: omarp@uaeh.edu.mx$^1$,
        ediazba@ipn.mx$^2$,
        lalopez@uaeh.edu.mx$^3$,
        dortizca@ipn.mx$^4$.}
	
	\begin{abstract}
		We investigate the statistical properties of the $q$-deformed Dirac oscillator for monolayer and bilayer graphene in the presence of a constant external magnetic field. Using $q$-deformed creation and annihilation operators, we obtain the corresponding energy spectra and eigenfunctions{\color{black}, as well as their probability current densities}, recovering the conventional Landau-level structures in the undeformed limit. For bilayer graphene, the lowest Landau levels are explicitly constructed within the $q$-deformed framework. The partition functions for both systems are derived using zeta-function techniques and the Euler–Maclaurin formula, allowing us to determine their main thermodynamic quantities for small values of the deformation parameter. Our results show that the $q$-deformation modifies the energy spectra{\color{black}, the behavior of the probability current density components,} and the thermodynamic response of the systems. A comparison between monolayer and bilayer graphene reveals distinct thermal behaviors, reflecting their different Landau-level structures and highlighting the influence of the $q$-deformation on their statistical properties.
		\\
		\\
		{\it Keywords:} Dirac oscillator, $q$-oscillator, statistical quantities, creation and annihilation operators, bilayer graphene.
	\end{abstract}
	
	\maketitle
	
	\section{Introduction}
	Among carbon allotropes, graphene continues to attract the attention of the theoretical scientific community as a novel low-dimensional material. Since its discovery, several studies have focused on the dynamics of monolayer graphene Dirac fermions under the interaction of external magnetic fields, leading to interesting phenomena such as the fractional quantum Hall effect or Landau-level quantization \cite{castro_guinea_2009}. Similarly, bilayer graphene has also been considered in the study of superconductivity \cite{McCann_2013}. For instance, rotated bilayer graphene at a magic angle $\theta\approx1.1^\circ$ exhibits a Moiré pattern, which opens the possibility of {\color{black}having superconductor structures} \cite{Cao2018}. {\color{black}On the other hand, s}ince Pacheco {\it et al.} \cite{PACHECO200393} studied the thermal properties of the one-dimensional Dirac oscillator for the first time, other authors have considered noncommutative spaces {\color{black}to describe the effect of noncommutative parameters on thermal quantities in both the two-dimensional Dirac oscillator} \cite{Boumali2013}, via the $\star$-Moyal product, and monolayer graphene under external magnetic fields \cite{Boumali_2015}. Similarly, phase-space noncommutativity has been investigated in the context of quantum cosmology (QC) \cite{Bastos2008} and black hole physics (BHP) \cite{Bastos2009}. In Ref. \cite{Bastos2013}, the Seiberg--Witten map was applied to monolayer graphene under an external magnetic field, yielding Landau levels modified by noncommutativity and showing that momentum noncommutativity implications are relevant not only in QC and BHP but also in graphene physics.\\
	\\
	{\color{black}Moreover}, $q$-deformation theory can be considered an alternative approach to viewing quantum mechanics as the limit of a more general theory that depends on one or more deformation parameters. Since the seminal works \cite{LCBiedenharn_1989,AJMacfarlane_1989}, several authors have applied the $q$-deformed Dirac oscillator to physical systems using the $q$-deformed oscillator algebra based on $q$-deformed creation and annihilation operators \cite{Eremin2006,HATAMI20163469,boumali_2017,Sargolzaeipo2018}. For instance, in Ref. \cite{houcca2021thermodynamic}, $q$-deformation theory is employed to obtain the thermodynamic properties of massless Dirac fermions in monolayer graphene with Rashba coupling. It is well known that several $q$-deformations of the same function have been proposed in different physical and mathematical contexts. In the present work, we employ the original form of $q$-deformation introduced by Heine in 1846 \cite{Heine1846}, who connected certain $q$-hypergeometric series with elliptic functions. A more recent formulation can be found in \cite{Eremin2006}. Other versions of $q$-deformation are discussed in Refs. \cite{79a378e8,Tsallis1988}.\\
	\\
	In this work, we consider the antisymmetric $q$-deformation theory as an approach to study the statistical properties of monolayer and bilayer graphene. The paper is organized as follows. In Sec. \ref{odm}, we review the basic concepts of the Dirac oscillator in monolayer graphene in the presence of a constant magnetic field{\color{black}, and also} we introduce the framework of the $q$-deformed Dirac oscillator for monolayer graphene. In Sec. \ref{zfmld}, we calculate {\color{black}the probability current density as well as} the partition function associated with the energy levels of the $q$-deformed Dirac oscillator in monolayer graphene using the Hurwitz zeta function and the residue theorem. In Sec. \ref{dong}, we discuss the Dirac oscillator in bilayer graphene and its $q$-deformed extension. Sec. \ref{zpbgqd} is devoted to 
	{\color{black}obtain the probability current density and} deriving the partition function of the Dirac oscillator in bilayer graphene using the Epstein--Hurwitz zeta function and the residue theorem. In Sec. \ref{sqt}, we calculate the thermodynamic quantities for both monolayer and bilayer graphene and analyze the main differences between the two systems. Finally, in Sec. \ref{scon}, we summarize the principal results and conclusions.
	
	\section{The Dirac Oscillator for Monolayer Graphene}\label{odm}
	Within the low-energy nearest-neighbor tight-binding approximation, the dynamics of massive Dirac fermions near the $\vec{K}$ and $\vec{K}'$ valleys in monolayer graphene is governed by the Hamiltonian:
	\begin{equation}\label{ec.hmg}
		{\bf H}_{\vec K}=v_{\rm F}
		\begin{pmatrix}
			0 & p_x-i p_y\\
			p_x+i p_y & 0
		\end{pmatrix}\,,
	\end{equation}
	where $v_{\rm F}$ denotes the Fermi velocity. The interaction between a charged quantum system and an external magnetic field is introduced through the minimal-coupling prescription, under which the momentum operator transforms as:
	\begin{equation}\label{key}
		{\bf p} \rightarrow {\bf p} + e\mathbf{A}(x)\,,
	\end{equation}
	where $\mathbf{A}(x)$ denotes the vector potential and $e$ is the charge of the particle. The corresponding magnetic field is obtained from the vector potential via $\mathbf{B}=\nabla\times\mathbf{A}$. 
	Since the vector potential is not uniquely defined, different gauge choices may be adopted. In the present work, we use the Landau gauge, $\mathbf{A}=(0,Bx,0)$, which corresponds to a uniform magnetic field $\mathbf{B}=(0,0,B)$. Therefore, the Hamiltonian in (\ref{ec.hmg}) is rewritten as:
	\begin{equation}\label{ec.hmg1}
		{\bf H}=v_{\rm F}
		\begin{pmatrix}
			0 & p_x-i p_y-ieBx\\
			p_x+i p_y +ieBx& 0
		\end{pmatrix}\,.
	\end{equation}
	The eigenvalue equation ${\bf H}\textcolor{black}
	{\bar{\Psi}(x,y)} = E\textcolor{black}
	{\bar{\Psi}(x,y)}$ can be simplified by exploiting the translational invariance of the system along the $y$-direction. Accordingly, the eigenfunction may be written as:
	\begin{equation}\label{ec}
		\textcolor{black}
		{\bar{\Psi}(x,y)}	=
		{\rm e}^{iky}\begin{pmatrix}
			\tilde{\psi}(x)\\
			\tilde{\phi}(x)
		\end{pmatrix}.
	\end{equation}
	Substituting the ansatz (\ref{ec}) into the eigenvalue equation yields the matrix representation:
	\begin{equation}\label{hamilto01}
		v_{\rm F}
		\begin{pmatrix}
			0 & p_x-ik\hbar-ieBx \\
			p_x+ik\hbar+ieBx & 0
		\end{pmatrix}
		\begin{pmatrix}
			\tilde{\psi}(x)\\
			\tilde{\phi}(x)
		\end{pmatrix}=E\begin{pmatrix}
			\tilde{\psi}(x)\\
			\tilde{\phi}(x)
		\end{pmatrix}\,.
	\end{equation}
	Introducing the change of variable $z=\sqrt{\omega/2}\left(x+2k/\omega\right)$, with $\omega=2eB/\hbar$, Eq. \eqref{hamilto01} can be rewritten as:
	\begin{equation}\label{hamilto02}
		v_{\rm F}\hbar\sqrt{\omega}
		\begin{pmatrix}
			0 & \frac{p_z}{\sqrt{2}\,\hbar}-i\frac{z}{\sqrt{2}} \\
			\frac{p_z}{\sqrt{2}\,\hbar}+i\frac{z}{\sqrt{2}} & 0
		\end{pmatrix}
		\begin{pmatrix}
			\psi(z)\\
			\phi(z)
		\end{pmatrix}=E\begin{pmatrix}
			\psi(z)\\
			\phi(z)
		\end{pmatrix}\,,
	\end{equation}
	{\color{black}where $\psi(z)=\tilde{\psi}(x)$ and $\phi(z)=\tilde{\phi}(x)$.} To solve the system algebraically, the Hamiltonian \eqref{hamilto02} is expressed in terms of Dirac creation and annihilation operators:
	\begin{eqnarray}\label{ops}
		{\bf a}_D^{\dagger}&=&\frac{p_z}{\sqrt{2}\,\hbar}+i\frac{z}{\sqrt{2}}\,,\\
		{\bf a}_D&=&\frac{p_z}{\sqrt{2}\,\hbar}-i\frac{z}{\sqrt{2}}\,,
	\end{eqnarray}
	which satisfy the commutation relations:
	\begin{equation}\label{comm}
		\left[{\bf a}_D,{\bf a}_D^{\dagger}\right]=1,\qquad
		\left[{\bf a}_D,{\bf a}_D\right]=0,\qquad
		\left[{\bf a}_D^{\dagger},{\bf a}_D^{\dagger}\right]=0\,.
	\end{equation}
	Finally, Eq. \eqref{hamilto02} takes the form:
	\begin{equation}\label{hamilto03}
		v_{\rm F}\hbar\sqrt{\omega}
		\begin{pmatrix}
			0 & {\bf a}_D \\
			{\bf a}_D^{\dagger} & 0
		\end{pmatrix}
		\begin{pmatrix}
			\psi(z)\\
			\phi(z)
		\end{pmatrix}=E\begin{pmatrix}
			\psi(z)\\
			\phi(z)
		\end{pmatrix}\,.
	\end{equation}
	
	\subsection{The $q$-Deformed Dirac Oscillator in Monolayer Graphene}\label{odmqd}
	
\begin{figure}[h]
		\centering
		\includegraphics[scale=1]{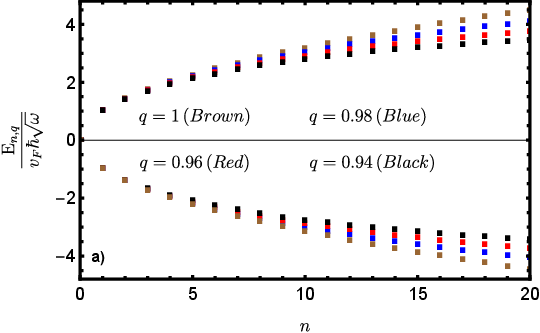}
		\includegraphics[scale=0.95]{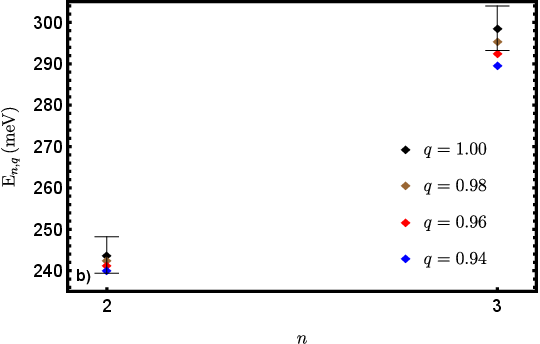}
		\caption{The left panel displays the dimensionless energy eigenvalues ${\rm E}_{n,q}$ given by Eq. (\ref{ecneq3}) for monolayer graphene as a function of the quantum number $n$, for several values of the $q$-deformation parameter. The right panel shows the second and third $q$-deformed Landau levels for different values of $q$, together with their associated uncertainties. The value $\omega=5.47\times10^{16}$ m$^{-2}$ was set.}
		\label{enm}
	\end{figure}

	The Dirac creation and annihilation operators define the $q$-deformed harmonic oscillator; these operators satisfy the relation:
	\begin{equation}\label{rec1}
		\left[{\bf a}_{D_q},{\bf a}_{D_q}^{\dagger}\right]={\bf a}_{D_q}{\bf a}_{D_q}^{\dagger}-q{\bf a}_{D_q}^{\dagger}{\bf a}_{D_q}=1\,,
	\end{equation}
	where the deformation parameter $q$ represents a real parameter in the range $0 < q < 1$, while ${\bf a}_{D_q}$ and ${\bf a}_{D_q}^{\dagger}$ represent the $q$-deformed annihilation and creation operators, respectively. 
	In the eigenbasis $\Psi_q\textcolor{black}(x,y)$, the Hamiltonian acts as:
	\begin{equation}
\mathbf{H}_q \Psi_q(x,y) = {\rm E}_{q} \Psi_q(x,y), \quad\mbox{with}\quad
\Psi_q(x,y)=e^{iky}\bar{\Psi}_q(x)\quad\mbox{and}\quad
\bar{\Psi}_q(x)=
\begin{pmatrix} \bar{\psi}_q(x) \\ \bar{\phi}_q(x) \end{pmatrix}.
\end{equation}
	Now, the Dirac equation (\ref{hamilto03}) for the $q$-deformed Hamiltonian reads:
	\begin{equation}\label{qhd1}
		v_{\rm F}\hbar\sqrt{\omega}
		\begin{pmatrix}
			0 & {\bf a}_{D_q} \\
			{\bf a}_{D_q}^{\dagger} & 0
		\end{pmatrix}
		\begin{pmatrix}
			\psi_q(z)\\
			\phi_q(z)
		\end{pmatrix}={\rm E}_q\begin{pmatrix}
			\psi_q(z)\\
			\phi_q(z)
		\end{pmatrix}\,.
	\end{equation}
	where $\bar{\psi}_q(x)=\psi_q(z)$ and $\bar{\phi}_q(x)=\phi_q(z)$ and $z$ is the change of variable used in the previous section. This system can be rewritten as the coupled equations:
	\begin{eqnarray}\label{eqkey1}
		v_{\rm F}\hbar\sqrt{\omega}{\bf a}_{D_q}\phi_q(z)&={\rm E}_q\psi_q(z)\,,\\\label{eqkey1b}
		v_{\rm F}\hbar\sqrt{\omega}{\bf a}_{D_q}^{\dagger}\psi_q(z)&={\rm E}_q\phi_q(z)\,.
	\end{eqnarray}
	Decoupling these equations yields:
	\begin{align}
		{\bf a}_{D_q}^{\dagger}{\bf a}_{D_q}\phi_q(z)&=\frac{{\rm E}^2_q}{v_{\rm F}^2\hbar^2\omega}
		\phi_q(z)\,,\label{key2}\\
        {\bf a}_{D_q}{\bf a}_{D_q}^{\dagger}\psi_q(z)&=\frac{{\rm E}^2_q}{v_{\rm F}^2\hbar^2\omega}\psi_q(z)\,.\label{key3}
	\end{align}
	The product ${\bf a}_{D_q}^{\dagger}{\bf a}_{D_q}$ is equal to the number operator $N_q$. Therefore, $\phi_q$ is an eigenfunction of this operator, with its corresponding eigenvalue given by the $q$-number $[n]$, which is defined by:
\begin{equation}\label{defn1}
		[n]=\frac{1-q^n}{1-q}\,.
\end{equation}
Similarly, given the commutation relation of the creation and annihilation operators in Eq.\eqref{rec1}, ${\bf a}_{D_q} {\bf a}_{D_q}^\dagger = 1 + q N_q$, which implies that the function $\psi_q$ is an eigenfunction of this operator. In this way, it is easy to verify that:
\begin{align}
		\left[v_{\rm F}^2\hbar^2\omega [n] - {\rm E}^2_{n,q}\right]\phi_{n,q}(z)&=0\,,\label{key3a}\\
        \left[v_{\rm F}^2\hbar^2\omega [n+1] - {\rm E}^2_{n,q}\right]\psi_{n,q}(z)&=0,\quad\mbox{for}\;n=0,1,...\label{key3b}
\end{align}
In this way, we can notice that $\phi_q$ possesses an additional energy level with respect to $\psi_q$. Consequently, the corresponding wave function solution takes the form $e^{iky} ((1-\delta_{n,0})\psi_{n-1,q}, \phi_{n,q})^T$. Therefore, the corresponding energy spectrum is given by:
	\begin{equation}\label{ecneq2}
		{\rm E}_{n,q}=\pm v_{\rm F}\hbar\sqrt{\omega}\sqrt{[n]}\,.
	\end{equation}
	Substituting $[n]$ from (\ref{defn1}) into (\ref{ecneq2}), we obtain the $q$-deformed Landau level energy spectrum:
	\begin{equation}\label{ecneq3}
		{\rm E}_{n,q}=\pm v_{\rm F}\hbar\sqrt{\omega}\sqrt{\frac{1-q^n}{1-q}}\,.
	\end{equation}
	Figure \ref{enm} a) presents the energy eigenvalues of $q$-deformed Dirac fermions in monolayer graphene. In the absence of deformation, the energy spectrum is quantized and exhibits a square-root dependence on the quantum number $n$. As deformation is introduced, the parabolic profile progressively narrows toward the $n$-axis. For fixed values of $n$, the energy levels approach one another regardless of the value of $q$ when $n$ is small; however, as $n$ increases, the effect of the deformation modifies these energy levels. 
	
	In the limit $q \to 1$, the energy spectrum reduces to the usual expression reported in Ref. \cite{castro_guinea_2009}:
	\begin{equation}\label{ec.llnq}
		E_n=\lim_{q\to1}{\rm E}_{n,q}=\pm v_{\rm F}\hbar\sqrt{\omega}\sqrt{n},\quad n=0,1,2,\dots\,.
	\end{equation}
	In Ref. \cite{PhysRevLett.98.197403}, infrared spectroscopic measurements under an external magnetic field enabled the experimental determination of the Landau-level spectrum in monolayer graphene. As indicated by Eq. (\ref{ec.llnq}), the transition energies between Landau levels scale with $\sqrt{B}\sim\sqrt{\omega}$. Measurements were carried out for magnetic fields up to $B=18\text{ T}$. Fitting the experimental data yielded the Fermi velocity:
	\begin{equation}
		v_{\rm F}=(1.12\pm0.02)\times10^6\text{ m/s}.
	\end{equation}
	Using this value of $v_{\rm F}$, the second and third $q$-deformed Landau levels obtained from Eq. (\ref{ecneq3}) are shown in Fig. \ref{enm} b) as a function of $n$, together with their associated uncertainties. To determine $q$, the energy of the undeformed level for $n=2$, $243.81$ meV, together with its experimental uncertainty of $\pm 4.4$ meV, was used as a reference. Taking the lower bound of this uncertainty range, the corresponding energy was $239.41$ meV. This value was then substituted into the energy expression of the $q$-deformed model given in Eq. (\ref{ecneq3}), and the parameter $q$ was determined numerically. This procedure yielded an approximate value of $q=0.928$.\\
	\\
    Now, to determine the wave functions, it is important to note that the system of equations \eqref{eqkey1} and \eqref{eqkey1b} indicates that $\psi_{n,q} \propto \mathbf{a}_{D_q} \phi_{n+1,q}$; therefore, it suffices to determine $\phi_{n,q}$. In this sense, let us begin by noting that for $n = 0$, the state $\phi_{0,q}$ is annihilated by $\mathbf{a}_{D_q}$. Furthermore, we introduce the coordinate representation of the Dirac creation and annihilation operators given in Ref.~\cite{Eremin2006}:
\begin{eqnarray}
    \mathbf{a}_{D_q} &=& \frac{\mathrm{e}^{-2i\alpha z} - \mathrm{e}^{i\alpha\frac{\mathrm{d}}{\mathrm{d}z}}\mathrm{e}^{-i\alpha z}}{-i\sqrt{1-\mathrm{e}^{-2\alpha^2}}} , \label{dpad} \\
    \mathbf{a}_{D_q}^{\dagger} &=& \frac{\mathrm{e}^{2i\alpha z} - \mathrm{e}^{i\alpha z}\mathrm{e}^{i\alpha\frac{\mathrm{d}}{\mathrm{d}z}}}{i\sqrt{1-\mathrm{e}^{-2\alpha^2}}} , \label{dpa1}
\end{eqnarray}
with $\alpha = \sqrt{-\ln q / 2}$ and $0 < \alpha < \infty$. This representation satisfies the relation given in equation \eqref{rec1} and recovers the standard harmonic oscillator expressions in the limit $q \to 1$ ($\alpha \to 0$). In this way,

\begin{equation}
\left( \mathrm{e}^{-2i\alpha z} - \mathrm{e}^{i\alpha\frac{\mathrm{d}}{\mathrm{d}z}}\mathrm{e}^{-i\alpha z} \right) \phi_{0,q}(z) = 0 ,
\end{equation}

\noindent which yields the normalized solution:

\begin{equation}
\phi_{0,q}(z) = \left(\frac{\pi\omega}{2}\right)^{\frac{1}{4}} {\rm e}^{-\frac{z^2}{2} + \frac{3i\alpha z}{2}} .
\end{equation}

On the other hand, to determine the wave functions associated with the excited states $\phi_{n,q}$, it is sufficient to consider the following relation:
\begin{equation}\label{eq29}
\phi_{n,q}(z)= \frac{\left(\mathbf{a}_{D_q}^{\dagger}\right)^n}{\sqrt{[n]!}} \, \phi_{0,q}(z), 
\end{equation}
with $[0]!=1$ and $[n]!=[1][2]...[n-1][n]$. Thus, the wave function of the $n$-th excited state is constructed by successive applications of the $q$-deformed Dirac creation operator $\mathbf{a}_{D_q}^\dagger$ on the ground state $\phi_{0,q}$:
\begin{equation}\label{eq30}
\phi_{n,q}(z) = N_n{\rm e}^{-\frac{z^2}{2} + \frac{3i\alpha z}{2}} \sum_{k=0}^{n} (-1)^k \frac{[n]!}{[k]![n-k]!} {\rm e}^{(n-k)2i\alpha z - k\alpha^2} ,
\end{equation}
where $N_n$ is a normalization constant given by:
\begin{equation}
    N_n = \frac{1}{\pi^{1/4} \sqrt{S_n(\alpha)}},   
\end{equation}
with $S_n(\alpha)$ being defined as:
\begin{equation}
S_n(\alpha) = \sum_{k=0}^{n} \sum_{j=0}^{n} (-1)^{k+j} \frac{([n]!)^2}{[k]![n-k]![j]![n-j]!} {\rm e}^{-\alpha^2 (k^2 + j^2 - 2kj + k + j)}.
\end{equation}
It is important to remark that in the limit $q \to 1$ ($\alpha \to 0$), the wave functions $\phi_{n,q}(z)$ recover the standard quantum harmonic oscillator solutions:

\begin{equation}
\phi_n(z) = \lim_{q \to 1} \phi_{n,q}(z) = \frac{1}{\sqrt{\pi 2^n n!}} H_n(z) {\rm e}^{-\frac{z^2}{2}} .
\end{equation}
	\section{Probability Current and partition Function for the $q$-Deformed Dirac Oscillator in Monolayer Graphene}\label{zfmld}
		
		\subsection{Monolayer Graphene Probability Current}
\begin{figure}[h]
		\centering
		\includegraphics[scale=0.9]{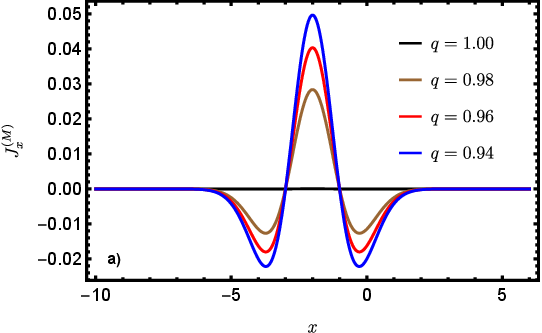}
		\includegraphics[scale=0.88]{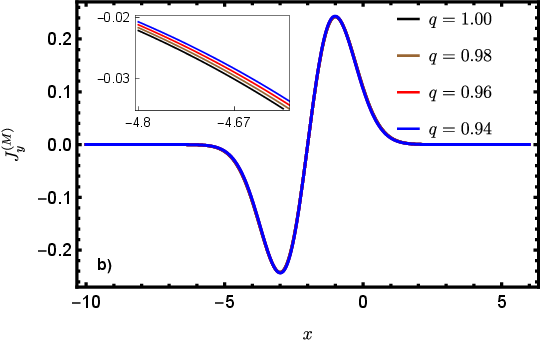}
		\caption{Spatial dependence of the probability current in monolayer graphene. 
			The left panel displays the $x$-component of the probability current, 
			$J_x^{(M)}$, for different values of the deformation parameter $q$, 
			whereas the right panel displays the $y$-component, $J_y^{(M)}$, 
			for the same values of $q$.  The values $k=1$ and $\omega=1$ were set.}
		\label{fig1}
	\end{figure}

		The probability current density $\mathbf{J}$ provides significant physical insights into the spatial dynamics and charge transport of relativistic quantum particles governed by Dirac-like equations. To evaluate the probability current for monolayer graphene within the $q$-deformed Dirac oscillator framework,  adopt the transformation $\psi(z) \to i\psi(z)$ for the lower spinor component in order to recover the standard probability current. \textcolor{black}{Thus, for these specific eigenstates, the spinorial wave function components lead to the components of the probability current vector $J^{(M)} = (J_x^{(M)}, J_y^{(M)})$:}
	\begin{eqnarray}
		J_x^{(M)} &=& 2v_{\text{F}} \, \text{Im}\!\left( \psi_{n-1,q}(z) \, \phi_{n,q}^*(z) \right), \\
		J_y^{(M)} &=&2v_{\text{F}} \, \text{Re}\!\left( \psi_{n-1,q}(z) \, \phi_{n,q}^*(z) \right).
	\end{eqnarray}
	In the undeformed limit ($q \to 1$), the $x-$component  probability current vanishes, $J_x^{(M)}=0$. However, introducing the $q$-deformation alters this behavior fundamentally. As a consequence of the $q$-deformed algebra and the underlying coordinate-space representation of the $q$-ladder operators, the spatial phase distribution is modified. Consequently, after applying the $q$-deformation, the probability current $J_x^{(M)}$ for excited states becomes non-zero. The non-vanishing behavior of $J_x^{(M)}$ as a function of the spatial coordinate $z$ and the deformation parameter $q$ is illustrated in Fig. \ref{fig1} a) and Fig. \ref{fig1} b), respectively. These modifications demonstrate that the $q$-deformation breaks the spatial current cancellation observed in the standard $q \to 1$ case, giving rise to persistent local probability currents in the excited states.
	
	\subsection{Partition function for monolayer graphene}
		
		On the other hand, to derive the thermodynamic properties of the $q$-deformed Dirac oscillator in monolayer graphene, we evaluate the partition function $Z$:
	\begin{equation}\label{ec.fpml}
		Z=\sum_{n=0}^{\infty}g_n {\rm e}^{-\beta {\rm E}_{n,q}},\quad \beta=\frac{1}{k_{\rm B} T},
	\end{equation}
	where $g_n$ denotes the degeneracy factor, $k_{\rm B}$ is the Boltzmann constant, and $T$ is the equilibrium temperature. Substituting the energy spectrum into this expression gives:
	\begin{equation}\label{key}
		Z=\sum_{n=0}^{\infty} {\rm e}^{-\tau \sqrt{[n]}},\quad \tau=\frac{\sqrt{2\hbar eB}\, v_{\rm F}}{k_{\rm B}T}=\frac{T_0}{T}\quad
		\mbox{and}\quad
		T_0=\sqrt{\frac{2e\hbar v^2_{\rm F}B}{k^2_{\rm B}}}.
	\end{equation}
	Here, the degeneracy factor is $g_n=1$ for all $n$. The parameter $T_0$ is a reference temperature that increases with the magnetic field, separating the high-temperature ($T \gg T_0$) and low-temperature ($T \ll T_0$) regimes. It plays a role analogous to the Debye temperature in solid-state physics \cite{Boumali2013}.\\
	\\
	To analyze small deformation effects, we set $q=1-\epsilon$ with $|\epsilon|\ll 1$. The partition function $Z$ can then be expanded as:
	\begin{equation}\label{fpml}
		Z\simeq\sum_{n=0}^{\infty}{\rm e}^{-\tau\sqrt{n}}\left(1+\frac{\tau\sqrt{n}(n-1)}{4}\epsilon\right)=Z_0+Z_1,
	\end{equation}
	where
	\begin{eqnarray}
		Z_0&=&\sum_{n=0}^{\infty}{\rm e}^{-\tau\sqrt{n}}=1+\sum_{n=0}^{\infty}{\rm e}^{-\tau\sqrt{n+1}},
		\label{fpmla}
		\\
		Z_1&=&\frac{\tau\epsilon}{4}\sum_{n=0}^{\infty}\sqrt{n}(n-1){\rm e}^{-\tau\sqrt{n}}.
		\label{fpmlb}
	\end{eqnarray}
	To evaluate $Z_0$, we follow Ref. \cite{Boumali_2015} and use the integral identity:
	\begin{equation}\label{ec.igml}
		{\rm e}^{-x}=\frac{1}{2\pi i}\int_Cx^{-s}\Gamma\left(s\right) {\rm d}s\,.
	\end{equation}
	The sum $Z_0$ becomes:
	\begin{equation}\label{key}
		Z_0=1+\frac{1}{2\pi i}\int_{C}{\rm d}s\, \tau^{-s} \sum_{n=0}^{\infty}(n+1)^{-\frac{s}{2}} \Gamma(s).
	\end{equation}
	Using the Hurwitz zeta function $\zeta_H(s,c)=\sum_{n=0}^{\infty}(n+c)^{-s}$, this is rewritten as:
	\begin{equation}
		Z_0=1+\frac{1}{2\pi i}\int_{C}{\rm d}s\, \tau^{-s} \zeta_{H}\left(\frac{s}{2},1\right) \Gamma(s). 
	\end{equation}
	Applying the residue theorem at $s=0$ and $s=2$, we obtain:
	\begin{equation}\label{ec.zoml}
		Z_0=1+\frac{\Gamma(2)}{\tau^2}+\zeta_{H}\left(0,1\right)=\frac{1}{2}+\frac{1}{\tau^2}.	
	\end{equation}
	The term $Z_1$ is evaluated using the Euler--Maclaurin formula:
	\begin{equation}\label{ec.em}
		\sum_{x=0}^{\infty} f(x)
		= \frac{1}{2}f(0)
		+ \int_{0}^{\infty} f(x)\,{\rm d}x
		- \sum_{p=1}^{\infty} \frac{B_{2p}}{(2p)!}\, f^{(2p-1)}(0)\,,
	\end{equation}
	where $B_{2p}$ are the Bernoulli numbers. Setting $f(x)=x\sqrt{x+1}\,{\rm e}^{-\tau\sqrt{x+1}}$, we find:
	\begin{equation}
		Z_1=\frac{\tau\epsilon}{4}\left[\frac{4(\tau+2)(\tau^2+3\tau+6)}{\tau^5}{\rm e}^{-\tau}
		-\sum_{p=1}^{\infty}\frac{B_{2p}}{(2p)!}f^{(2p-1)}(0)\right].
	\end{equation}
	For $T>T_0$, the leading contribution corresponds to $p=1$:
	\begin{equation}
		Z_1=\epsilon\left[\frac{1}{\tau}\left(1+\frac{2}{\tau}\right)\left(1+\frac{3}{\tau}+\frac{6}{\tau^2}\right)-\frac{\tau}{28}\right]{\rm e}^{-\tau}.\label{ec.z1mlz}
	\end{equation}
	Combining Eqs. (\ref{ec.zoml}) and (\ref{ec.z1mlz}), the $q$-deformed partition function for monolayer graphene is:
	\begin{equation}\label{eczpmg}
		Z=\frac{1}{2}+\frac{1}{\tau^2}+\epsilon\left[\frac{1}{\tau}
		+\frac{5}{\tau^2}
		+\frac{12}{\tau^3}
		+\frac{12}{\tau^4}-\frac{\tau}{28}\right]{\rm e}^{-\tau}.
	\end{equation}
	
	\section{The Dirac Oscillator for Bilayer Graphene}\label{dong}
	Similarly to \eqref{ec.hmg}, the effective Hamiltonian near $\vec{K}$ for bilayer graphene is:
	\begin{equation}\label{ec.hbg}
		{\bf H}_{\vec K}=\frac{1}{2m^{*}}
		\begin{pmatrix}
			0 & (p_x-i p_y)^2 \\
			(p_x+i p_y)^2 & 0
		\end{pmatrix}\,.
	\end{equation}
	In the presence of a perpendicular magnetic field along $x$, minimal coupling modifies the momentum via ${\bf p} \rightarrow {\bf p} + e\mathbf{A}(x)$. In the Landau gauge $\mathbf{A}=(0,Bx,0)$, the eigenvalue problem ${\bf H}\bar{\Psi}(x,y)=E\bar{\Psi}(x,y)$ leads to:
	\begin{equation}\label{hamilto}
		{\bf H}\bar{\Psi}(x,y)=\frac{1}{2m^{*}}
		\begin{pmatrix}
			0 & \Pi^{2} \\
			(\Pi^{\dagger})^{2} & 0
		\end{pmatrix}
		\bar{\Psi}(x,y)=E\bar{\Psi}(x,y)\,,
	\end{equation}
	where $\Pi = p_x - i p_y - ieA(x)$. Exploiting translational invariance along $y$, we use the ansatz:
	\begin{equation}\label{psi}
		\bar{\Psi}(x,y)=        
        {\rm e}^{ik y}
        \Psi(x),\quad\mbox{with}\quad\Psi(x)=
		\begin{pmatrix}
			\tilde{\psi}(x)\\
			\tilde{\phi}(x)
		\end{pmatrix}\,.
	\end{equation}
	Substituting \eqref{psi} into \eqref{hamilto} yields:
	\begin{equation}\label{hamilto1}
		\frac{1}{2m^{*}}
		\begin{pmatrix}
			0 & \left(p_x-ieBx-ik\hbar\right)^2 \\
			\left(p_x+ieBx+ik\hbar\right)^2 & 0
		\end{pmatrix}
		\begin{pmatrix}
			\tilde{\psi}(x)\\
			\tilde{\phi}(x)
		\end{pmatrix}=E\begin{pmatrix}
			\tilde{\psi}(x)\\
			\tilde{\phi}(x)
		\end{pmatrix}\,.
	\end{equation}
	Under the change of variable $z=\sqrt{\omega/2}\left(x+2k/\omega\right)$, {\color{black}$\psi(z)=\tilde{\psi}(x)$ and $\phi(z)=\tilde{\phi}(x)$, as before}, Eq. \eqref{hamilto1} becomes:
	\begin{equation}\label{hamilto2}
		\frac{\hbar^2\omega}{2m^{*}}
		\begin{pmatrix}
			0 & \left(\frac{p_z}{\sqrt{2}\,\hbar}-i\frac{z}{\sqrt{2}}\right)^2 \\
			\left(\frac{p_z}{\sqrt{2}\,\hbar}+i\frac{z}{\sqrt{2}}\right)^2 & 0
		\end{pmatrix}
		\begin{pmatrix}
			\psi(z)\\
			\phi(z)
		\end{pmatrix}=E\begin{pmatrix}
			\psi(z)\\
			\phi(z)
		\end{pmatrix}\,,
	\end{equation}
	or in terms of Dirac creation and annihilation operators:
	\begin{equation}\label{hamilto3}
		\frac{\hbar^2\omega}{2m^{*}}
		\begin{pmatrix}
			0 & {\bf a}_D{\bf a}_D \\
			{\bf a}_D^{\dagger}{\bf a}_D^{\dagger} & 0
		\end{pmatrix}
		\begin{pmatrix}
			\psi(z)\\
			\phi(z)
		\end{pmatrix}=E\begin{pmatrix}
			\psi(z)\\
			\phi(z)
		\end{pmatrix}\,.
	\end{equation}
	
	\subsection{The $q$-Deformed Dirac Oscillator in Bilayer Graphene}\label{dongqd}

This offers a unified representation of all eigenstates. Then, equation \eqref{hamilto3} for the $q$-deformed Hamiltonian reads:
\begin{equation}\label{qhd}
    \frac{\hbar^2\omega}{2m^{*}}
    \begin{pmatrix}
        0 & {\mathbf{a}_{D_q}}{\mathbf{a}_{D_q}} \\
        {\mathbf{a}_{D_q}}^{\dagger}{\mathbf{a}_{D_q}}^{\dagger} & 0
    \end{pmatrix}
    \begin{pmatrix}
        \psi_q(z)\\
        \phi_q(z)
    \end{pmatrix} = \text{E}_q
    \begin{pmatrix}
        \psi_q(z)\\
        \phi_q(z)
    \end{pmatrix},
\end{equation}
leading to the coupled equations:
\begin{align}\label{key1}
    \frac{\hbar^2\omega}{2m^{*}}{\mathbf{a}_{D_q}}{\mathbf{a}_{D_q}}\phi_q(z) &= \text{E}_q\psi_q(z)\,,\\
    \frac{\hbar^2\omega}{2m^{*}}{\mathbf{a}_{D_q}}^{\dagger}{\mathbf{a}_{D_q}}^{\dagger}\psi_q(z) &= \text{E}_q\phi_q(z)\,.
\end{align}
Decoupling for $\phi_q(z)$ and $\psi_q(z)$ yields:
\begin{align}
    \label{key2}
    {\mathbf{a}_{D_q}}^{\dagger}{\mathbf{a}_{D_q}}^{\dagger}{\mathbf{a}_{D_q}}{\mathbf{a}_{D_q}}\phi_q(z) - \frac{4{m^{*}}^2}{\hbar^4\omega^2}\text{E}^2_q\phi_q(z) &= 0\,,\\
    {\mathbf{a}_{D_q}}{\mathbf{a}_{D_q}}{\mathbf{a}_{D_q}}^{\dagger}{\mathbf{a}_{D_q}}^{\dagger}\psi_q(z) - \frac{4{m^{*}}^2}{\hbar^4\omega^2}\text{E}^2_q\psi_q(z) &= 0.
\end{align}
From the commutation relation, it can be easily demonstrated that:
\begin{equation}
    {\mathbf{a}_{D_q}}^\dagger {\mathbf{a}_{D_q}} = \frac{{\mathbf{a}_{D_q}} {\mathbf{a}_{D_q}}^\dagger - 1}{q},
\end{equation}
therefore:
\begin{align}
   {\mathbf{a}_{D_q}}^\dagger {\mathbf{a}_{D_q}}^\dagger \,{\mathbf{a}_{D_q}} {\mathbf{a}_{D_q}} 
    &= \frac{{\mathbf{a}_{D_q}}^\dagger {\mathbf{a}_{D_q}} {\mathbf{a}_{D_q}}^\dagger {\mathbf{a}_{D_q}} - {\mathbf{a}_{D_q}}^\dagger {\mathbf{a}_{D_q}}}{q} \\
    &= \frac{N_q [N_q - 1]}{q}.
\end{align}
Thus, the following equation must be satisfied:
\begin{equation}
    \left[ \frac{[n]}{q} ([n] - 1) - \frac{4{m^{*}}^2 {\rm E}_q^2}{\hbar^4 \omega^2} \right]\phi_{n,q}(z) = 0.
\end{equation}
Given the definition for $[n]$ in Eq. \eqref{defn1}, we obtain:
\begin{equation}
    \frac{[n]}{q} ([n] - 1) = [n][n-1],
\end{equation}
which leads to the energy eigenvalues:
\begin{equation}
    {\rm E}_{n,q} = \pm \frac{\hbar^2 \omega}{2m^{*}} \sqrt{[n][n-1]},
\end{equation}
where the eigenfunctions $\phi_{n,q}$ are defined by the relations in Eqs.~\eqref{eq29} and \eqref{eq30}. 

Similarly, it is straightforward to show that for $\psi_{n,q}$, the eigenfunctions are also defined by the same equations, but their energy levels are given by:
\begin{equation}
    {\rm E}_{n,q} = \pm \frac{\hbar^2 \omega}{2m^{*}} \sqrt{[n+1][n+2]}.
\end{equation}
That is, $\phi_{n,q}$ possesses two additional levels in comparison with $\psi_{n,q}$. Beyond this detail, note that $n$ is not the ideal label for the energy levels, as $n=0,1$ actually correspond to degenerate states of the system. Therefore, it is convenient to perform the following restructuring using a new label that properly tracks the state counting, as described below:

\textcolor{black}{In order to include both the two degenerate ground states and the excited states, we define the spinor eigenstates as:
\begin{equation}
    \Psi_{m,q}(x) =
    \begin{pmatrix}
        (1-\delta_{n,0})(1-\delta_{n,1})\,\psi_{n-2,q}(z)\\[2mm]
        \phi_{n,q}(z)
    \end{pmatrix},
\end{equation}
where $\psi_n(z)$ and $\phi_n(z)$ are the $q$-deformed harmonic-oscillator eigenfunctions in Eqs.~\eqref{eq29} and \eqref{eq30}, and $\delta_{i,j}$ denotes the Kronecker delta. The spinor index $m$ is related to $n$ by:
\begin{equation}
    m = (1-\delta_{n,0})(1-\delta_{n,1})(n-1).
    \label{53}
\end{equation}
For $n=0$ and $n=1$ (ground states), the upper component vanishes and $m=0$:
\begin{equation}
    \Psi_{0,q}^{(1)}(x)=
    \begin{pmatrix}
        0\\
        \phi_{0,q}(z)
    \end{pmatrix},
    \qquad
    \Psi_{0,q}^{(2)}(x)=
    \begin{pmatrix}
        0\\
        \phi_{1,q}(z)
    \end{pmatrix}.
\end{equation}
For excited states ($n \ge 2$), $m = n-1$, and the spinors simplify to:
\begin{equation}
    \Psi_{n-1,q}(x)=
    \begin{pmatrix}
        \psi_{n-2,q}(z)\\[2mm]
        \phi_{n,q}(z)
    \end{pmatrix},
    \qquad n \ge 2,
\end{equation}}
where their energy spectrum is given by:
\begin{equation}\label{ecneq1}
    \text{E}_{m,q} = \pm\frac{\hbar^2\omega}{2m^{*}}\sqrt{[n][n-1]}\,.
\end{equation}
The full $q$-deformed Landau level spectrum for bilayer graphene is:
\begin{equation}\label{ecneq}
    \text{E}_{m,q} = \pm\frac{\hbar^2\omega}{2m^{*}}
    \sqrt{\left(\frac{1-q^n}{1-q}\right)\left(\frac{1-q^{n-1}}{1-q}\right)}\,,
\end{equation}
which, in the limit $q \to 1$, satisfies \cite{Fernandez}:
\begin{equation}\label{key3}
    E_m = \lim_{q\to1}\text{E}_{m,q} = \pm\frac{\hbar^2\omega}{2m^{*}}\sqrt{n(n-1)},\quad n=1,2,\dots\,.
\end{equation}

\begin{figure}[h]
    \centering
    \includegraphics[scale=1]{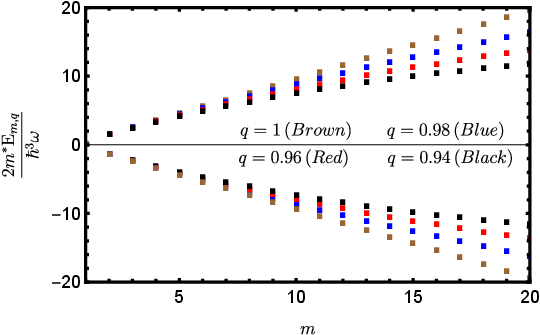}
    \caption{Energy eigenvalues ${\rm E}_{m,q}$ \eqref{ecneq} in dimensionless units versus quantum number $m$ for bilayer graphene at different values of $q$.}
    \label{enb}
\end{figure}

Figure~\ref{enb} shows the dimensionless energy eigenvalues of $q$-deformed Dirac fermions in bilayer graphene. As deformation increases, the curves narrow toward the $m$-axis, exhibiting a trend similar to that seen in monolayer graphene.

	\color{black}
	
	\section{Probability Current and partition Function for the $q$-Deformed Dirac Oscillator in Bilayer Graphene}\label{zpbgqd}
	
	\subsection{Bilayer Graphene Probability Current}
	
	For bilayer graphene, the effective low-energy Hamiltonian yields a probability current operator involving spatial derivatives of the spinor components. The components of the probability current vector $\mathbf{J}^{(B)} = (J_x^{(B)}, J_y^{(B)})$ are expressed{\color{black}, respectively,} as:
	\begin{equation}
		J_x^{(B)} = \frac{\hbar}{m^*} \sqrt{\frac{\omega}{2}}\, \text{Im}\!\left[ \psi_{n-1,q}^*(z) \left( \frac{{\rm d}\phi_{n,q}(z)}{{\rm d}z} \right) + \phi_{n,q}^*(z) \left( \frac{{\rm d}\psi_{n-1,q}(z)}{{\rm d}z} \right)+2z\psi_{n-1,q}^*(z)\phi_{n,q}(z) \right],
	\end{equation}
	\begin{equation}
		J_y^{(B)} = \frac{\hbar}{m^*}\sqrt{\frac{\omega}{2}}\, \text{Re}\!\left[ \phi_{n,q}^*(z) \left( \frac{{\rm d}\psi_{n-1,q}(z)}{{\rm d}z} \right) - \psi_{n-1,q}^*(z) \left( \frac{{\rm d}\phi_{n,q}(z)}{{\rm d}z} \right)-2z\psi_{n-1,q}^*(z)\phi_{n,q}(z) \right].
	\end{equation}
	
\begin{figure}[h]
		\centering
		\includegraphics[scale=0.9]{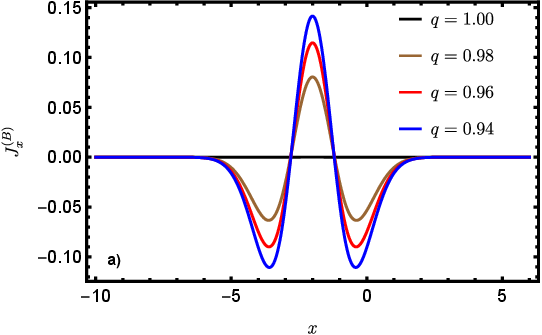}
		\includegraphics[scale=0.88]{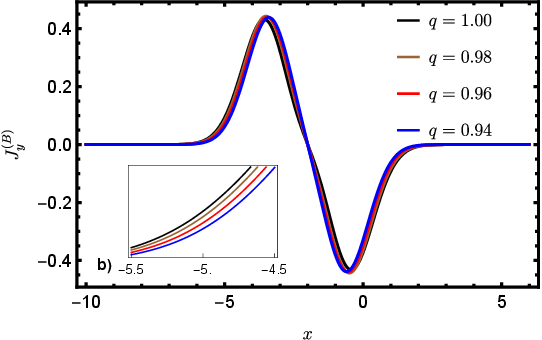}
		\caption{Spatial dependence of the probability current in bilayer graphene.
			The left panel displays the $x$-component of the probability current, $J_x^{(B)}$, whereas the right panel displays the corresponding $y$-component, $J_y^{(B)}$, for different values of the deformation parameter $q$. The values $k=1$ and $\omega=1$ were set.}
		\label{fig3}
	\end{figure}

	As in the monolayer case, the implementation of the $q$-deformation leads to noticeable modifications in the probability current profiles compared to the undeformed case ($q \to 1$). The resulting spatial distributions for $J_x^{(B)}$ and $J_y^{(B)}$ in bilayer graphene across different deformation levels are presented in Fig. \ref{fig3} a) and Fig. \ref{fig3} b). These figures confirm that the $q$-deformed algebraic structure introduces shifts in the local current amplitudes and alters the spatial node locations, providing further evidence of how algebraic deformations impact physical observables in low-dimensional carbon allotropes.
	
	\subsection{Partition function for bilayer graphene}
	
	Again, using the energy spectrum in Eq. (\ref{ecneq}), the corresponding partition function $Z$ at temperature $T$ is:
	\begin{equation}\label{ec.fp}
		Z=\sum_{m=0}^{\infty}g_m {\rm e}^{-\beta {\rm E}_{m,q}}\quad\longrightarrow\quad Z=2+\sum_{n=2}^{\infty}{\rm e}^{-\tau\sqrt{\left(\frac{1-q^n}{1-q}\right)\left(\frac{1-q^{n-1}}{1-q}\right)}},
	\end{equation}
	where $g_0=2$ and $g_m=1$ for $m\geq 2$. Defining $\tau=\frac{\hbar^2\omega\beta}{2m^{*}}=\frac{T_0}{T}$ with $T_0=\frac{\hbar^2\omega}{2m^{*}k_{\rm B}}$, the positive-energy partition function is:
	\begin{equation}\label{key}
		Z^{B}=\sum_{n=1}^{\infty} g_n{\rm e}^{-\tau\sqrt{[n][n-1]}}\,.
	\end{equation}
	Expanding up to first order in $\epsilon = 1-q$:
	\begin{eqnarray}
		Z^{B}
		&\simeq& 2+\sum_{n=2}^{\infty} {\rm e}^{-\tau\sqrt{n\left(n-1\right)}}
		+\frac{\epsilon\tau}{4}\sum_{n=1}^{\infty} g_n\sqrt{n\left(n-1\right)}\left(2n-3\right) {\rm e}^{-\tau\sqrt{n\left(n-1\right)}}\nonumber\,,\\
		&=& 2+Z_0^{B} +\frac{\epsilon\tau}{4}Z_1^{B}+\frac{\epsilon\tau}{2}\frac{\partial}{\partial\tau}Z_0^{B}\,, \label{ecfpbc}
	\end{eqnarray}
	with the auxiliary functions:
	\begin{eqnarray}
		Z_0^{B}&=&\sum_{n=0}^{\infty}{\rm e}^{-\tau\sqrt{\left(n+\frac{3}{2}\right)^2-\frac{1}{4}}}\,,\label{z0aux}\\
		Z_1^{B}&=&\sum_{n=0}^{\infty}\sqrt{\left(n+1\right)\left(n+2\right)}\left(2n+3\right){\rm e}^{-\tau\sqrt{\left(n+1\right)\left(n+2\right)}}\,.\label{z1aux}
	\end{eqnarray}
	Using the integral identity (\ref{ec.igml}), $Z_0^{B}$ is written as:
	\begin{equation}\label{ecu.zb0}
		Z_0^{B}=\frac{1}{2\pi i}\int_C {\rm d}s\,\tau^{-s} \sum_{n=0}^{\infty}\left[\left(n+\frac{3}{2}\right)^2-\frac{1}{4}\right]^{-\frac{s}{2}}\Gamma\left(s\right)\,.
	\end{equation}
	Using the inhomogeneous Epstein--Hurwitz zeta function $E_1^{c}(s; a_1; c_1)$ \cite{elizalde1994zeta} and evaluating the poles at $s=0$ and $s=1$, we obtain:
	\begin{equation}\label{fpz0a}
		Z_0^{B}=\frac{1}{2}\zeta_H\left(0,\frac{3}{2}\right)+\frac{1}{2\tau}=-\frac{1}{2}+\frac{1}{2\tau}\,.
	\end{equation}
	Evaluating $Z_1^{B}$ via the Euler--Maclaurin formula for $T>T_0$ ($\tau<1$) gives:
	\begin{equation}\label{ec.z1a}
		Z_1^{B}=\epsilon\left[1+\frac{\sqrt{2}}{\tau}+\frac{1}{\tau^2} +\frac{55\sqrt{2}}{192}\tau\right]{\rm e}^{-\sqrt{2}\,\tau}\,.
	\end{equation}
	Thus, the total partition function for bilayer graphene is:
	\begin{equation}\label{fpbi}
		Z^{B}=\frac{3}{2}+\frac{1}{2\tau}+
		\epsilon\left[\frac{1}{4\tau}+\left(1+\frac{\sqrt{2}}{\tau}+\frac{1}{\tau^2} +\frac{55\sqrt{2}}{192}\tau\right){\rm e}^{-\sqrt{2}\,\tau}\right]\,.
	\end{equation}
	
	\section{Thermodynamic Properties of the $q$-Deformed Dirac Oscillator}\label{sqt}
	
	From the results obtained in the previous section, we observe that the partition functions for monolayer graphene, given by (\ref{eczpmg}), and bilayer graphene, given by (\ref{fpbi}), exhibit different temperature dependences owing to their distinct energy spectra. In monolayer graphene, the Landau-level spectrum depends on the square root of the quantum number, whereas the effective low-energy spectrum of bilayer graphene has a different structure. Consequently, the two systems have different densities of thermally accessible states and therefore display distinct thermal behaviors. In particular, the partition function of monolayer graphene (\ref{eczpmg}), increases more rapidly with the dimensionless temperature $\tau^{-1}=T/T_0$, whereas the partition function of bilayer graphene (\ref{fpbi}), increases at a slower rate.
		
		Once the partition functions for monolayer and bilayer graphene have been obtained, the main thermodynamic quantities, including the Helmholtz free energy, entropy, internal energy, and specific heat, can be derived from the following relations:
		\begin{eqnarray}
			F^i&=&-\frac{a_i}{\tau}\ln Z^i\label{ec.fbi}\,,
			\\
			U^i&=&-a_i\frac{\partial}{\partial\tau}\ln Z^i\,,\label{ec.ubi}\\
			S^i&=&\frac{k_{\rm B}\tau^2}{a_i}\frac{\partial F^i}{\partial \tau}\,,\label{ec.sbi}\\
			C^i&=&-k_{\rm B}\frac{\tau^2}{a_i}\frac{\partial U^i}{\partial \tau}	\,,\label{ec.cbi}
		\end{eqnarray}
		Here, the index $i$ labels the graphene system under consideration, with $i=M$ corresponding to monolayer graphene and $i=B$ corresponding to bilayer graphene. The parameters $a_i$ are defined as $a_M=\sqrt{2e\hbar v_{\rm F}^2B}$ and $a_B=(\hbar^2\omega)/2m^*$.
	
	\begin{figure}[h]
		\centering
		\includegraphics[scale=0.95]{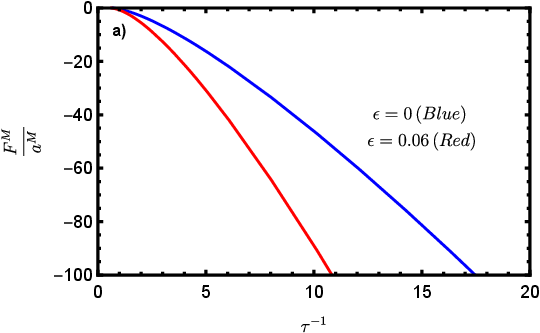}
		\includegraphics[scale=0.95]{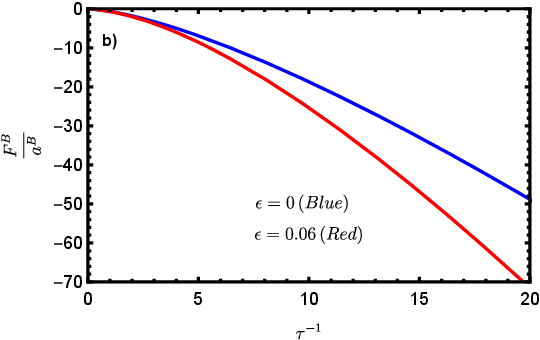}
		\caption{Free energy of the $q$-deformed Dirac oscillator versus dimensionless temperature $\tau^{-1}$ for $\epsilon=0$ and $\epsilon=0.06$: (a) monolayer graphene, (b) bilayer graphene.}
		\label{fe}
	\end{figure}
	
	\begin{figure}[h]
		\centering
		\includegraphics[scale=0.9]{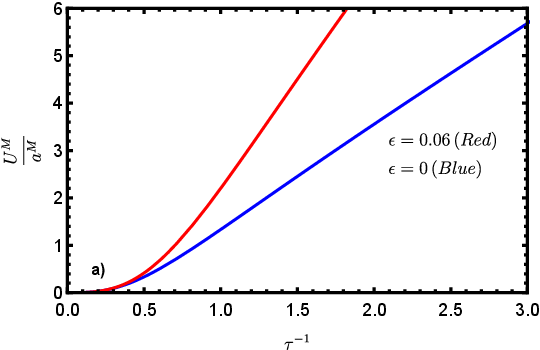}
		\includegraphics[scale=0.9]{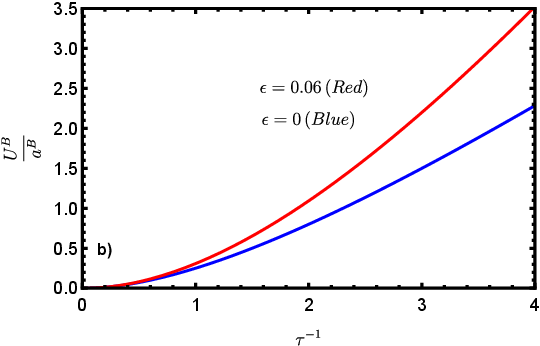}
		\caption{Internal energy versus dimensionless temperature $\tau^{-1}$ for $\epsilon=0$ and $\epsilon=0.06$: (a) monolayer graphene, (b) bilayer graphene.}
		\label{ue}
	\end{figure}
	
	\begin{figure}[h]
		\centering
		\includegraphics[scale=0.85]{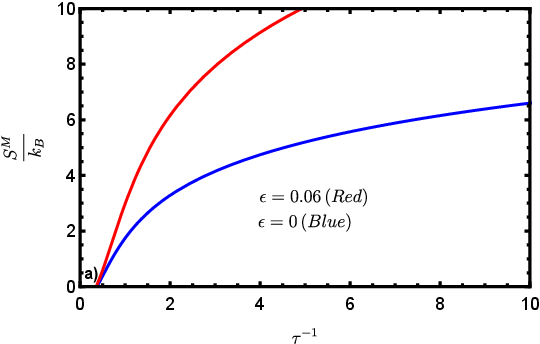}
		\includegraphics[scale=0.85]{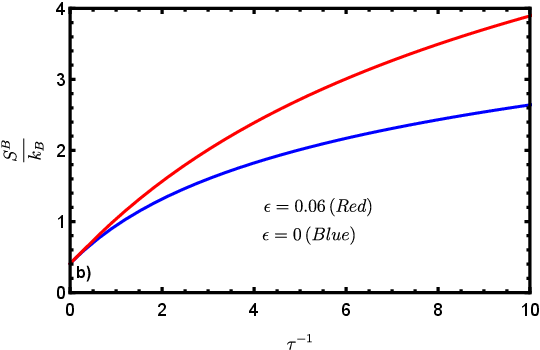}
		\caption{Entropy versus dimensionless temperature $\tau^{-1}$ for $\epsilon=0$ and $\epsilon=0.06$: (a) monolayer graphene, (b) bilayer graphene.}
		\label{se}
	\end{figure}
	
	\begin{figure}[h]
		\centering
		\includegraphics[scale=0.85]{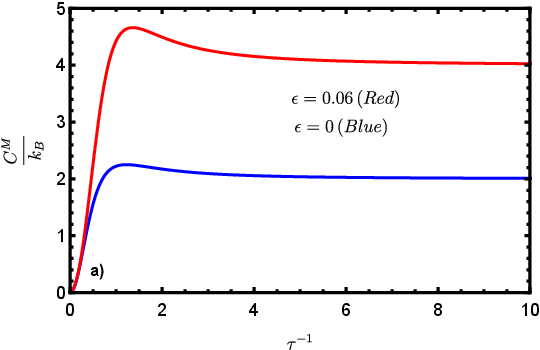}
		\includegraphics[scale=0.85]{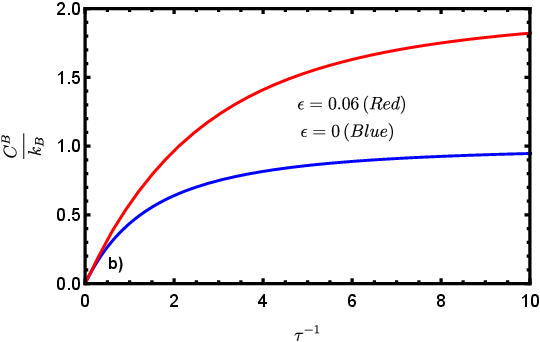}
		\caption{Specific heat versus dimensionless temperature $\tau^{-1}$ for $\epsilon=0$ and $\epsilon=0.06$: (a) monolayer graphene, (b) bilayer graphene.}
		\label{ce}
	\end{figure}
	
	The thermodynamic quantities are plotted in Figures \ref{fe}-\ref{ce}. All quantities were computed as functions of the dimensionless temperature $\tau^{-1}$ and the deformation parameter $\epsilon$. To illustrate the influence of $\epsilon$ on the thermodynamic functions, we set {\color{black}$\epsilon=0.06$} in all the plots. The results show that the effects of the $q$-deformation vanish in the limit $\epsilon\to0$, thereby recovering the undeformed case.
		
		Figure \ref{fe} shows the Helmholtz free energy as a function of the dimensionless temperature $\tau^{-1}$. As expected, the free energy decreases with increasing temperature. Moreover, this decrease becomes more pronounced for larger values of $\epsilon$ in both monolayer and bilayer graphene.
		
		Figure \ref{ue} shows that the internal energy increases with temperature. Furthermore, for both monolayer and bilayer graphene, the internal energy increases as $\epsilon$ increases.
		
		As shown in Figure \ref{se}, the entropy increases smoothly with both temperature and $\epsilon$, without exhibiting any abrupt changes in either graphene system. The absence of such discontinuities suggests that the curvature observed in the specific-heat curves in Figure \ref{ce} is not associated with a phase transition in either monolayer or bilayer graphene.
		
		At high temperatures, the specific heat approaches a constant value for both monolayer and bilayer graphene. In the undeformed  limit,  $C^{M} \rightarrow 2k_{\rm B}$ and  $C^{B} \rightarrow k_{\rm B}$, respectively. This saturation behavior is analogous to the high-temperature Dulong--Petit regime. A similar behavior  was previously reported for $q$-deformed massless Dirac fermions in monolayer graphene with Rashba coupling \cite{houcca2021thermodynamic}, where the undeformed system approaches $C = 2k_{\rm B}$, while the introduction of the deformation increases the asymptotic heat capacity. In the present model,  the $q$-deformation also enhances the thermal response, although the limiting values differ between monolayer and bilayer graphene. This difference could be attributed to their distinct Landau-level structures and, consequently, to the different number and distribution of thermally accessible effective degrees of freedom.
		
		In general, the internal-energy, entropy, and specific-heat curves for monolayer graphene lie above the corresponding curves for bilayer graphene. In contrast, the Helmholtz free energy exhibits the opposite behavior.
		
		Finally, we expect that, once experimental measurements of the thermodynamic properties of monolayer and bilayer graphene under an external magnetic field become available, our results will provide a useful theoretical framework for investigating these properties within the $q$-deformed Dirac oscillator model.
	
	\section{Conclusion}\label{scon}

	In this work, we derived the energy spectra of the Dirac oscillator in the presence of an external magnetic field within the framework of $q$-deformation for both monolayer and bilayer graphene. We showed how the corresponding energy eigenvalues and eigenfunctions can be obtained within the proposed deformation formalism{\color{black}, and how the corresponding probability current densities are modified. After applying a $q$-deformation to the quantum system, the probability current components are significantly modified. In particular, the deformation induces non-zero current components. Nevertheless, in the limit $q \to 1$, the standard unperturbed behavior is smoothly recovered, confirming the consistency of the deformed model.} We also calculated the partition functions and, from them, the main thermodynamic quantities, including the Helmholtz free energy, internal energy, entropy, and specific heat, for both graphene systems.
		
		The results reveal clear differences between the thermodynamic responses of monolayer and bilayer graphene. The larger internal energy and entropy obtained for monolayer graphene could indicate that its energy spectrum allows a stronger thermal population of accessible states within the temperature range considered.
		
		A particularly interesting result concerns the high-temperature behavior of the specific heat. In the undeformed limit, the specific heat approaches constant values for both monolayer and bilayer graphene, displaying a Dulong–Petit-like saturation. The introduction of the $q$-deformation enhances the thermal response in both systems. However, the magnitude of this enhancement differs between monolayer and bilayer graphene, reflecting their distinct Landau-level spectra and thermally accessible degrees of freedom.
		
		Finally, we expect that, once high-precision experimental measurements of the thermodynamic properties of graphene become available, our results may provide a useful theoretical framework for investigating these properties.\\
        \\
        \noindent \textbf{Acknowledgments}
The authors acknowledge the Secretaría de Ciencia, Humanidades, Tecnología e Innovación (SECIHTI) for the support provided during the development of this work. EDB also acknowledges the SIP-IPN research grant 20260933.\\
\\
\textbf{Declaration of generative AI and AI-assisted technologies in the manuscript preparation process.} AI was used to check references and perform typesetting checks. The author is responsible for the paper's mathematical content.
	\bibliographystyle{unsrt}
	\bibliography{bibliografia}
\end{document}